# Demo Abstract: WiSwitch: A Low-Cost WiFi-based Remote Switch Control for Smart Homes

Yazeed Alrubyli
Prince Sultan University, Saudi Arabia
212110966@psu.edu.sa

Anis Koubaa
Prince Sultan University, Saudi Arabia
CISTER/INESC-TEC, Polytechnic Institute of Porto, Portugal
akoubaa@coins-lab.org

## Abstract

**With the emergence of the IoT, there is an increasing trend towards designing new low-cost cyber-physical platforms accessible through the Internet. In this Demo paper, we present WiSwitch, a home-made prototype of a low-cost wireless switch control allowing the user to control the room light through the cloud. The hardware is based on commercial-off-the-shell (COTS) ESP 12 WiFi module with basic electronic circuitry. We develop an embedded client application in the ESP 12 that allows to connect it to Amazon Web Services cloud and update the status of the switch remotely. We also present an experimental validation of the WiSwitch platform.**

## 1 Introduction

Internet of Things (IoT) is driving the IT world towards an unprecedented integration of physical systems with the cyber world and namely the Internet. Everyday, new IoT platforms are being released opening new insights for different kinds of innovative applications, including smart home, smart environment, precision agriculture, security and surveillance, and several others. Gartner predicts that the value add of the IoT is expected to reach $1.9 Trillion **[REF]** by 2020, while CISCO and Ericsson postulates that the number of IoT devices will reach 50 billion by 2020. In particular, smart home applications are considered as number one in the top 10 most popular applications of the IoT [1] as of December 2015. There has been an increasing trend of manufacturing smart home devices that are connected to the Internet. Cost-effectiveness and interoperability are the two most important design requirements considering: (1) the large-scale deployment of these devices, (2) the need to interoperate with heterogeneous like smart phone, web applications and others.

In this paper, we contribute with the design of WiSwitch, a low-cost remote light switch that addresses the aforementioned requirements. First, it uses COTS low-cost hardware and mainly based on the ESP12e WiFi chip with an overall cost that does not exceed $5. Second, we propose a cloud-based solution to ensure the interoperability of the WiSwitch platform with any mobile device or Web application. We leverage the use of the Web server embedded in the ESP12e WiFi chip to establish the connectivity with the cloud and thus with the end-user through the cloud. We use Amazon Web Services cloud to illustrate our prototype, although it can be deployed on any other cloud platform. In addition, we opted for the use of WiFi rather than other wireless solution (like IEEE 802.15.4 and Bluetooth) because it is Internet-friendly, and is available on commodity devices like smart phones and tablets. In addition, in this particular case we do not have stringent energy constraints as the WiSwitch platform is mains-powered (110$V$-220$V$ supported) like typical light switches.

The motivation behind the WiSwitch is to provide an easy control and monitoring of light switches in smart home environments to promote life comfort.

## 2 Hardware Architecture

Figure 1 illustrates the WiSwitch hardware platform. The platform consists of the ESP-12e WiFi module, a voltage regulator, and one channel relay. ESP-12 is the heart of the WiSwitch platform, and consists of a WiFi chip (IEEE 802.11 b/g/n) and a built-in 32-bit micro-controller. Furthermore, the regulator is used to adapt the voltage from 110$V$/220$V$ to 5$V$ in order to power the ESP12e WiFi module and the channel relay. Finally, the channel relay is used to switch the light ON/OFF based on the signal received from the ESP12e WiFi module.

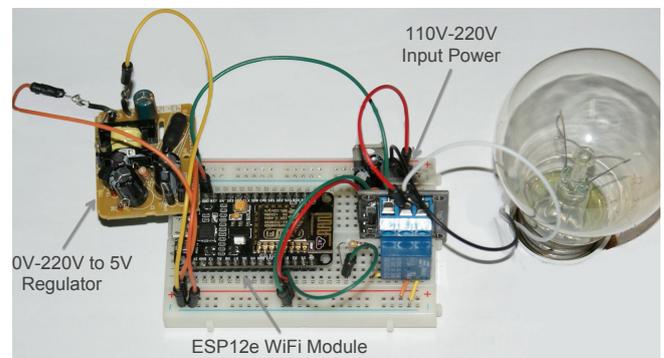

**Figure 1. WiSwitch hardware platform**

The added value of this simple platform is its ability to connect to a cloud server, transmit the status of the light switch whether it is ON or OFF, and also receive control command from the client application through the cloud server.

---

[1] http://iot-analytics.com/10-internet-of-things-applications/

Table 1. WiSwitch Hardware Characteristics

| Categories | Items | Values |
|---|---|---|
| WiFi Paramters | WiFi Protocles | 802.11 b/g/n |
| | Frequency Range | 2.4G-2.5G (2400M-2483.5M) |
| Hardware Paramaters | Operating Voltage | 3.0 3.6V |
| Software Parameters | WiFi mode | station/softAP/SoftAP+station |
| | Security | WPA/WPA2 |
| | Network Protocols | IPv4, TCP/UDP/HTTP/FTP |
| | User Configuration | AT Instruction Set, Cloud Server, Android/iOS App |

More details on the software architecture will be presented in the next section.

The design of the circuit is focuses on reducing the complexity and the cost of to reach consumers at the lowest price as standard switch or socket. The estimated costs of WiSwitch platform is around $5 ($2 for the ESP12e WiFi module, $2 Voltage Regulator, and $1 for one channel relay). The use of WiFi wireless transeiver has more benefits as compared to Bluetooth or RFID approaches which are more limited in terms of coverage and Internet support.

Table 1 summarizes the characteristics of the WiSwitch platform.

## 3 System Architecture

This system architecture consists of three software layers:

- **The IoT Layer**: it represents the software suite of the WiSwitch platform. Using the Arduino IDE of the ESP12e WiFi chip, it is programmed using Lua programming language to read a JSON message received from the cloud containing whenever there is a status change. The exchanged JSON messages are (`Switch:on` or `Switch:off`) and the the relay switched accordingly to either ON or OFF, respectively. The (ArduinoJson) library is used to handle the JSON processing of JSON message received from the cloud in order to decide sending a on/off signal to the relay.

- **The Cloud Layer**: We used Amazon Web Services to host the server application that interfaces with the client application. The cloud service receives the commands from the client application, which can be deployed as a smartphone application or standard Web application, and forward them to the switch relay. In addition, the cloud store every command or light status in a database for future analysis of all light control operations. This help to post-process the log information stored in the database to estimate the total operation time of a particular light switch or also to detect possible threats, like for example turning ON the light by a non-authorized intruder when the house residents are outside. In addition, the cloud can be pre-programmed to send alarms in case of threat detection.

- **The Client Layer**: This layer consists of client applications used to control the WiSwitch platform. The use of the platform-independent JSON format allows to deploy the application on any kind of device, as such coping with device heterogeneity and ensuring interoperability.

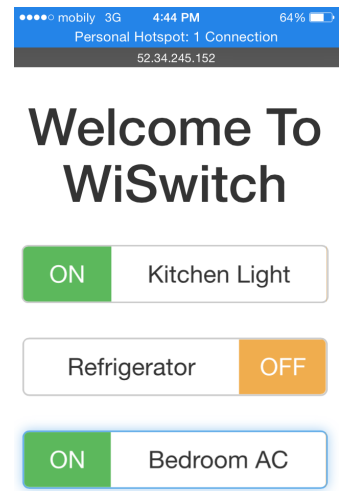

Figure 2. WiSwitch hardware platform

In our current deployment, we used a web application to interact with WiSwitch platform through the cloud. The client has two connection options. In case of Internet connectivity available, the client application can connect directly the server on the cloud to monitor or control the switch. In case of non availability of Internet connection, the client can still monitor and control the WiSwitch platform in local mode through a local area network without interfacing with the cloud. This dual mode operation provide better flexibility to users whether they are inside or outside the local network of the WiSwitch. Figure 2 presents the client interface on a mobile phone.

In the Demo, we intend to demonstrate the deployment of this system and illustrate its benefits.

## 4 Discussion and Ongoing Work

We are working towards extending the WiSwitch system to incorporate smart functionalities such as pre-programming the switches in a smart building, auto-detecting threats in case of a non expected switch on or off. In addition, we aim at developing REST web services to interact with the WiSwitch.